# Dense Coding Capacity in Correlated Noisy Channels with Weak Measurement

Jin-Kai Li (李进开), Kai Xu (徐凯)，Guo-Feng Zhang (张国锋)*

*School of Physics, Beihang University, Beijing 100191, China;*

**Abstract**：Capacity of dense coding via correlated noisy channel is greater than that in uncorrelated noisy channel. It is shown that weak measurement and reversal measurement can make further effort to improve quantum dense coding capacity in correlated amplitude damping channel, but this effort is very small in correlated phase damping channel and correlated depolarizing channel.

## I. INTRODUCTION

Since the concept of quantum dense coding (sometimes called superdense coding) is originally introduced by Bennett CH and Wiesner SJ [1], this field has been discussed widely. To realize this communication protocol, it is essential to share an entangled state between the sender (Alice) and receiver (Bob) initially. The entangled state has the property that can be transformed by the sender into another state via a local operation, taken from some set of operations. Then Alice transmits her qubit to Bob, who performs an measurement on the global state together with the received qubit and his original one. The signal state that Alice sent is distinguished unambiguously by the measurement. Thus, sending a single qubit transmits 2 bits of classical information. This is absolutely impossible without entanglement; the amount of information conveyed by an isolated qubit cannot exceed one bit. Mattle et al[2] have experimentally realized quantum dense coding in optical experiments with polarization-entangled photons.

It's straightforward to be concerned with the amount of classical information that can be reliably transmitted by quantum states, i.e., the capacity for dense coding, subsequently. It has been proved that, for noiseless channels and unitary encoding, the dense coding capacity is given by

$$\chi = S(\bar{\rho}) - S(\rho_0), \quad (1)$$

where $\rho_0$ is the initial resource state shared between Alice and Bob, $\bar{\rho}$ represents the density matrix after quantum dense coding, $S$ is the von Neumann entropy and $\chi$ is the capacity of quantum dense coding. For the density matrix $\rho_0$, the von Neumann entropy is written explicitly: $S(\rho_0) = -\sum_Z \lambda_Z \log_2 \lambda_Z$, where $\lambda_Z$ are eigenvalues of the density matrix $\rho_0$. Recent decades, attention has been paid to many scenarios of super dense coding over noiseless channels[3]-[4]. Not only the maximally entangled state but more general initial state of the two particles has been discussed by Barenco and Ekert[5]. More general case for higher-dimensional[7] entanglement have been argued, which provides more capacity of quantum dense coding over conventional qubit entanglement. Dense coding for continuous variables has been argued by Braunstein et al[8].

---

* Corresponding author. gf1978zhang@buaa.edu.cn

In addition, the case of uncorrelated noisy channel (i.e., memoryless channel) has also been discussed[9]. Because noise is unavoidably present in reality, optical fibers and an unmodulated spin chain[10] are practicle applications of such quantum noisy channels which are appropriate for long- and short-distance quantum communication, respectively. Physically, noise is a process that arises through interaction with the environment. Mathematically, a noisy quantun channel can be described as a completely positive trace preserving (CPTP) linear map $\Lambda$, acting on the quantum state. For uncorrelated cases (channels and states) where the von Neumann entropy fulfills a specific condition, the superdense coding capacity was derived.

The above studies discussed the sequence of qubits passing through an uncorrelated channel by neglecting the correlations between multiple uses of quantum channels. However, correlation effects of the quantum channel can't neglect along with the transmission rate raising in quantum channel, as it has been practically explored in the solid-state implementation of fiber[11] or quantum hardware that suffers from low-frequency noise[12]. Quantum correlated channels have thus caused a lot of attention recently.

In this paper, we consider the quantum dense coding in correlated noisy channels. Three common noise sources are taken into account: the amplitude damping channel, phase damping channel and depolarizing channel. Additionally, we analyze the effects of weak measurement and reversal measurement on the capacity of dense coding. The particular plan without weak measurement and reversal measurement is that the initial state directly pass through the correlated channel and then we make a dense coding. As an improvement on the former, the plan with weak measurement and reversal measurement is "weak measurement + correlated noise channel+ reversal measurement+ dense coding".

This paper is organized as follows. In Sec.Ⅱ we briefly introduce the correlated quantum channel and the quantum dense coding capacity. Section Ⅲ discusses the influence of correlation strength of the channel and the initial entanglement on the capacity of quantum dense coding for different noisy models. The effects of weak measurement and reversal measurement on capacity are also discussed in this section. The conclusions drawn from the present study are given in Sec. Ⅳ.

## II. CORRELATED QUANTUM CHANNEL AND OPTIMAL DENSE CODING CAPACITY

Correlated noisy channel is the further approximation to actual transmission of memoryless noisy channel[13]. Let us consider general memoryless noise channel first. The effect of transmission channels is described by Kraus operators $E_k$, satisfing $\sum_k E_k^\dagger E_k = 1$. That means sending one qubit in density operator $\rho_0$ through the channel the output state is determined by the map $\rho_0 \to \Lambda(\rho_0) = \sum_k E_k \rho_0 E_k^\dagger$. It's helpful to think about the example of Pauli channels in which the Kraus operators to individual qubit can be described by the Pauli operators $\sigma_{x,y,z}$, i.e. $E_k = \sqrt{p_k}\sigma_k$, with $\sum_k p_k = 1$, $k = 0, x, y, z$ and $\sigma_0 = I$, i.e., the identity operator of second order, where $p_x, p_y, p_z$ is the probability of the rotation of an angle $\pi$ around the corresponding axes $\hat{x}, \hat{y}, \hat{z}$ on the qubit state and $p_0$ is the probability of identity. A channel is memoryless when its effect on arbitrary state $\rho$, consisting of n qubits (including entangled ones), is described by $\Lambda(\rho) = \sum_{k_1 \cdots k_n} \left( E_{k_n} \otimes \cdots \otimes E_{k_1} \right) \rho \left( E_{k_1}^\dagger \otimes \cdots \otimes E_{k_n}^\dagger \right)$. Pauli channels is given by Kraus operators of following form: $E_{k_1 \cdots k_n} = \sqrt{p_{k_1 \cdots k_n}} \sigma_{k_1} \cdots \sigma_{k_n}$, with $\sum_{k_1 \cdots k_n} p_{k_1 \cdots k_n} = 1$, $k_i = 0, x, y, z$. The quantity $p_{k_1 \cdots k_n}$ can be understood as the probability that a given random sequence of rotations of an angle $\pi$ along axes $k_1 \cdots k_n$ is applied to the sequence of n qubits sent through the channel. For a memoryless channel, $p_{k_1 \cdots k_n} = p_{k_1} p_{k_2|k_1} \cdots p_{k_n|k_{n-1}}$, where $p_{k_n|k_{n-1}}$ can be interpreted as the conditional probability that a $\pi$ rotation around the $k_n$ axis is applied to the nth qubit given that a $\pi$ rotation around the $k_{n-1}$ axis was applied on the (n-1)th qubit. Here we will consider the case of two consecutive uses of a channel with partial memory, i.e., we will assume

$$p_{k_n|k_{n-1}} = (1-\mu) p_{k_n} + \mu \delta_{k_n k_{n-1}}. \tag{2}$$

This means that after the (n-1)th qubit rotates an angle $\pi$ along axis $k_{n-1}$, the nth qubit acts the same rotation with probability $\mu$ or rotates an angle $\pi$ along axis $k_n$ with probability $(1-\mu) p_{k_n}$. Thus, $\mu \in [0,1]$ can be understood as the degree of classical correlation in the channel. When $\mu = 0$, the model depicts uncorrelated channel, while for $\mu = 1$, the model depicts fully correlated channel.

This so-called correlated noise channel can describe situations in which time correlations can't neglect in the system. For example, $\mu$ is a function of the time lapse between the two channel uses. If two qubits sent almost at the same time, the properties of the channel, which controls the rotation, will be unchanged, and it is rational to assume that the operators on both qubits will be in the manner $E_k^c = \sqrt{p_k}\sigma_k \sigma_k$, where the superscript "c" means correlated, which means two qubits perform the same rotation of an angle $\pi$ along axis $k$ with probability $p_k$. It is easy to check that the Bell states $|\Psi_\pm\rangle = \frac{1}{\sqrt{2}}\{|01\rangle \pm |10\rangle\}, |\Phi_\pm\rangle = \frac{1}{\sqrt{2}}\{|00\rangle \pm |11\rangle\}$ are eigenstates of the operators $E_k^c$, where $|0\rangle$ and $|1\rangle$ are eigenstates of the operator $\sigma_z$. As a result, they will pass undisturbed via the correlated Pauli noisy channel.

If the time interval between the channel uses is so long that the channel properties have changed, then the operator will be $E_{k_1 k_2}^u = \sqrt{p_{k_1}}\sqrt{p_{k_2}}\sigma_{k_1}\sigma_{k_2}$, where the superscript "$u$" means uncorrelated, which means the first qubit rotates an angle $\pi$ along axis $k_1$ with probability $p_{k_1}$ while the second qubit rotates $\pi$ along axis $k_2$ with probability $p_{k_2}$.

An intermediate case is described in the form $E_{k_1 k_2}^i = \sqrt{p_{k_1}\left[(1-\mu)p_{k_2} + \mu\delta_{k_1 k_2}\right]}\sigma_{k_1}\sigma_{k_2}$, where the superscript "$i$" means intermediate, which means after the first qubit rotates $\pi$ along axis $k_1$ with probability $p_{k_1}$, the second qubit acts the same rotation with probability $\mu$ or rotates an angle $\pi$ along axis $k_2$ with probability $(1-\mu)p_{k_2}$. The Kraus operator for two consecutive uses of a general partial correlated channel, not only Pauli channels, is in the form

$$E_{k_1 k_2} = \sqrt{p_{k_1}[(1-\mu)p_{k_2} + \mu\delta_{k_1 k_2}]}B_{k_1} \otimes B_{k_2}. \tag{3}$$

where $B_{k_1}$ and $B_{k_2}$ are the Kraus operators for a single qubit of the corresponding channel.

With the above description, we can get the final state of the system through a correlated noise channel

$$\rho_1 = (1-\mu)\sum_{k_1 k_2} E_{k_1 k_2}\rho_0 E_{k_1 k_2}^\dagger + \mu\sum_k E_{kk}\rho_0 E_{kk}^\dagger. \tag{4}$$

where $E_{k_1 k_2}$ represents the Kraus operator in uncorrelated noise channel and $E_{kk}$ represents that in the correlated noise channel.

Then we introduce the capacity for dense coding. Ref.[14] has shown that the $d^2$ signal states $(i_{max} = d^2 - 1)$ generated by mutually orthogonal unitary transformations with equal probabilities yield the maximum $\chi$. The set of mutually orthogonal unitary transformations of dense coding for two qubits are

$$\begin{aligned} U_{00}|x\rangle &= |x\rangle, \\ U_{10}|x\rangle &= e^{i\pi x}|x\rangle, \\ U_{01}|x\rangle &= |x+1(\bmod 2)\rangle, \\ U_{11}|x\rangle &= e^{i\pi x}|x+1(\bmod 2)\rangle, \end{aligned} \tag{5}$$

where $|x\rangle$ is the single qubit computational basis. The average state of the ensemble of signal states generated by above equation is

$$\bar{\rho} = \frac{1}{4}\sum_{0}^{3}(U_i \otimes I)\rho(U_i^\dagger \otimes I), \quad (6)$$

where $I$ is the identity matrix of second order and $\rho$ represents the density matrix that needs to make dense coding. For simplicity, we suppose $0 \to 00; 1 \to 01; 2 \to 10; 3 \to 11$.

## III. CAPACITY IN THREE CORRELATED NOISE CHANNELS

The capacity of dense coding through correlated noisy channel is given by

$$\chi = S(\bar{\rho}_1) - S(\rho_1), \quad (7)$$

where $\rho_1$ is given by Eq.(4) and $\bar{\rho}_1$ is given by Eq.(6). The initial entangled state has changed after going through the correlated noisy channel. So the density matrices that we use are different from Eq (1). Below we mainly discuss the above problems in three common correlated noise channels: amplitude damping channel, phase damping channel and depolarizing channel.

**Capacity in correlated amplitude damping channel**

The amplitude damping (AD) channel can also be used to describe the spontaneous emission of a photon by a two-level atom at low or zero temperature[15][16]. The Kraus operators for one qubit are

$$B_{k_1} = \begin{pmatrix} 1 & 0 \\ 0 & \sqrt{\lambda} \end{pmatrix}, \quad B_{k_2} = \begin{pmatrix} 0 & \sqrt{1-\lambda} \\ 0 & 0 \end{pmatrix}, \quad (8)$$

where $\lambda = e^{-\Gamma t}$ is the decay of the excited population and $\Gamma$ is the dissipation rate. Two qubits transmitting through the partial correlated AD channel can be described by Eq. (4). We should pay attention that the Kraus operators $E_{k_1 k_2}$ in uncorrelated AD channel for two qubits transmission are considered in the following form:

$$E_{k_1 k_2} = B_{k_1} \otimes B_{k_2}, \quad (k_1, k_2 = 0, 1), \quad (9)$$

while the Kraus operators $E_{kk}$ in full-memory amplitude damping channel introduced in Ref.[17], different from $E_{k_1 k_2}$ are given by

$$E_{00} = \begin{pmatrix} 1 & 0 & 0 & 0 \\ 0 & 1 & 0 & 0 \\ 0 & 0 & 1 & 0 \\ 0 & 0 & 0 & \sqrt{\lambda} \end{pmatrix}, \quad E_{11} = \begin{pmatrix} 0 & 0 & 0 & \sqrt{1-\lambda} \\ 0 & 0 & 0 & 0 \\ 0 & 0 & 0 & 0 \\ 0 & 0 & 0 & 0 \end{pmatrix}. \quad (10)$$

In this paper, we take into account the initial state $\rho_0 = |\Phi\rangle\langle\Phi|$, where $\Phi = \alpha|00\rangle + \beta|11\rangle$ corresponds to the Bell-like state with $0 \leq \alpha \leq 1$. Since this communication protocol is based on quantum entanglement, let $\alpha$ and $\beta$ be $\sqrt{2}/2$ in discussion in this paper so as to take advantage of maximum entanglement. According to Eq.(4), in the correlated AD channel, the evolved density matrix $\rho_1$ of the two-qubit system, whose elements in the standard computational basis $\{|1\rangle = |11\rangle, |2\rangle = |10\rangle, |3\rangle = |01\rangle, |4\rangle = |00\rangle\}$ are

$$\begin{aligned}
\rho_{11} &= \alpha^2 - (-1+\lambda)\beta^2\left[1+\lambda(-1+\mu)\right], \\
\rho_{22} &= (-1+\lambda)\lambda\beta^2(-1+\mu), \\
\rho_{33} &= (-1+\lambda)\lambda\beta^2(-1+\mu), \\
\rho_{44} &= \lambda\beta^2(\lambda+\mu-\lambda\mu), \\
\rho_{14} &= \rho_{41} = -\lambda\alpha\beta(-1+\mu) + \sqrt{\lambda}\alpha\beta\mu.
\end{aligned} \qquad (11)$$

Based on Eq. (6), we can obtain the density matrix of $\bar{\rho}_1$. Then according to Eq. (7), the capacity of quantum dense coding can be obtained analytically, but it is too long to write here. By confirming the capacity $\chi$, the influences of the correlation strength of channel $\mu$ on capacity are depicted in Fig.1. The blue line with square in Fig.1 corresponds to $\mu = 0$, which means the correlation strength of channel is zero, and what's more, it corresponds to the uncorrelated channel. The purple line with circle, the yellow line with rhombus, the green line with triangle and the light blue with inverted triangle represent to $\mu = 0.25, 0.5, 0.75, 1$, respectively. From this figure, we can reasonably infer that the capacity of quantum dense coding increases monotonically with the increasing correlation strength $\mu$ when the damping coefficient $\lambda$ of AD channel is fixed. So we can conclude that the correlated channel increases the capacity of quantum dense coding comparing to the uncorrelated channel.

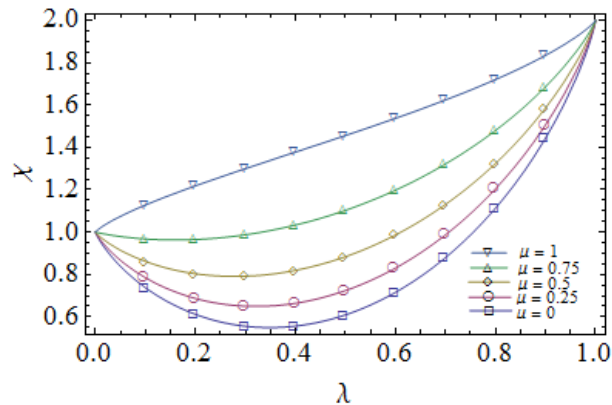

Fig.1. The dependence of the capacity of quantum dense coding on the damping coefficient

$\lambda$ in correlated amplitude damping channel, where $\alpha = \beta = \sqrt{2}/2$. The blue line with square, purple line with circle, yellow line with rhombus, green line with triangle and light blue line with inverted triangle (from bottom to top) in plot correspond to the correlation strength of channel $\mu = 0, 0.25, 0.5, 0.75, 1$, respectively.

From Ref.[18], we know weak measurement and reversal measurement improve the capacity of dense coding. So we consider the influence of weak measurement on capacity in correlated noise channel.

We know standard quantum measuring procedure make an initial state collapse to an eigenstate of the observable. Different from standard measurement, weak measurement[19] make so little influence on initial state that almost no initial state collapse to its eigenstate. Weak measurement and reversal measurement have been studied theoretically[20] and experimentally[21].

Weak measurement operator and reversal measurement operator for two qubits can be written as:

$$M_w(m_1, m_2) = \begin{pmatrix} 1 & 0 \\ 0 & \sqrt{1-m_1} \end{pmatrix} \otimes \begin{pmatrix} 1 & 0 \\ 0 & \sqrt{1-m_2} \end{pmatrix},$$
$$M_{rev}(n_1, n_2) = \begin{pmatrix} \sqrt{1-n_1} & 0 \\ 0 & 1 \end{pmatrix} \otimes \begin{pmatrix} \sqrt{1-n_2} & 0 \\ 0 & 1 \end{pmatrix}. \tag{12}$$

$M_w(m_1, m_2)$ is a weak measurement operator, $m_1$ and $m_2$ are weak measurement strengths. In the same presentation, $M_{rev}(n_1, n_2)$ is the reversal measurement operator, $n_1$ and $n_2$ are reversal measurement strengths. For simplicity, we assumed that $m_1 = m_2 = m$ and $n_1 = n_2 = n$.

The capacity of quantum dense coding with weak measurement and reversal measurement is

$$\chi = S(\bar{\rho}_2) - S(\rho_2), \tag{13}$$

where $\rho_2$ is the density matrix in the "with weak measurement and reversal measurement" plan and $\bar{\rho}_2$ is that matrix after dense coding given by Eq.(6). The density matrix can be described by

$$\rho_2 = \frac{1}{T} \begin{pmatrix} \rho_{11} & 0 & 0 & \rho_{14} \\ 0 & \rho_{22} & 0 & 0 \\ 0 & 0 & \rho_{33} & 0 \\ \rho_{41} & 0 & 0 & \rho_{44} \end{pmatrix}, \tag{14}$$

where $T$ is the normalized factor. In this equation:

$$\begin{aligned}
\rho_{11} &= (-1+n)^2(\alpha^2 + (-1+\lambda)(-1+m)^2\beta^2(-1+\lambda-\lambda\mu)) \\
\rho_{22} &= \rho_{33} = -(-1+\lambda)\lambda(-1+m)^2(-1+n)\beta^2(-1+\mu) \\
\rho_{44} &= \lambda(\beta-m\beta)^2(\lambda+\mu-\lambda\mu) \\
\rho_{14} &= \rho_{41} = \sqrt{\lambda}(1-m)(1-n)\alpha\beta(-\sqrt{\lambda}(-1+\mu)+\mu).
\end{aligned} \quad (15)$$

According to Eq.(13), the capacity $\chi$ of quantum dense coding with weak measurement and reversal measurement is shown in Fig.2. As for the research of the capacity of quantum dense coding under the influence of weak measurement and reversal measurement, we learn that no matter how the damping coefficient $\lambda$ changes that maximum value can be got by adjusting weak measurement strength and reversal measurement strength. For the purpose of simplicity, we assume that the damping coefficient $\lambda$ is a certainty and then discuss the effects of weak measurement and reversal measurement. To make it easier to understand, we use a point to analyze. For example, the capacity of quantum dense coding with weak measurement and reversal measurement is 1.7494 for weak measurement strength $m=0.9$ and reversal measurement strength $n=0.95$ when the damping coefficient of AD channel $\lambda=0.5$ and the correlation strength of channel $\mu=0.5$. Meanwhile, the capacity of quantum dense coding without weak measurement and reversal measurement is 0.8842. In this case, the capacity of quantum dense coding under the weak measurement and reversal measurement is greater than that without weak measurement and reversal measurement. Furthermore, the method of weak measurement and reversal measurement can be used for different damping coefficient $\lambda$, which can make dense coding successful and improve the capacity of dense coding in AD channel.

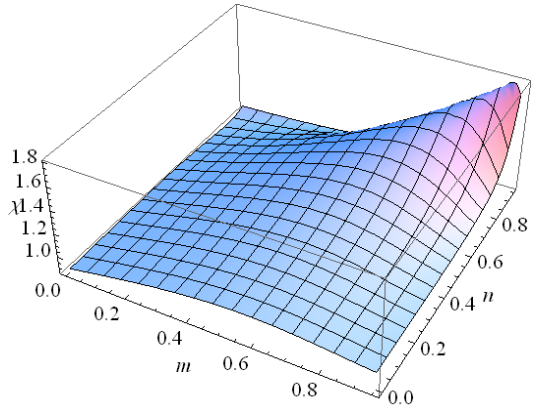

Fig.2 The capacity of quantum dense coding is plotted as function of weak measurement strength $m$ and reversal measurement strength $n$ when the damping coefficient $\lambda$ is 0.5 and correlation strength $\mu$ is 0.5, where $\alpha=\beta=\sqrt{2}/2$.

## Capacity in correlated phase damping channel

The phase damping channel describes a decoherencing process without exchanging energy with the environment. The Kraus operators of a single qubit in the phase damped channel are represented as the Pauli operators $\sqrt{p_0}\sigma_0$ and $\sqrt{p_z}\sigma_z$. As we discussed in Sec. II, the Kraus operators for two qubits passing through a memoryless phase damping channel can be written as

$$E_{k_1 k_2} = \sqrt{p_{k_1} p_{k_2}} \sigma_{k_1} \otimes \sigma_{k_2}, \quad (16)$$

where $k_1, k_2 = (0, z)$, $p_0 = (1+\lambda)/2$, $p_z = (1-\lambda)/2$ and $\lambda = \exp(-\Gamma t)$. For the partial correlated phase damping channel, the Kraus operators $E_{kk}$ are given as

$$E_{kk} = \sqrt{p_k}\sigma_k \otimes \sigma_k, \quad (k = 0, z). \tag{17}$$

Substituting Eq. (14) and (15) into Eq. (4), the density matrix elements of the two qubits in the correlated phase damped channel can be expressed in the following form:

$$\begin{aligned}\rho_{11} &= \alpha^2, \\ \rho_{22} &= \beta^2, \\ \rho_{14} &= \rho_{41} = \alpha\beta\left(-\lambda^2(-1+\mu)+\mu\right).\end{aligned} \tag{18}$$

Based on Eq. (7), the capacity of quantum dense coding can be obtained analytically. The influences of the correlation strength of channel $\mu$ on capacity are depicted in Fig.3. The blue line with square in Fig.3 corresponds to $\mu = 0$, which corresponds to the uncorrelated channel. The purple line with circle, the yellow line with rhombus, the green line with triangle and the light blue line with inverted triangle represent to $\mu = 0.25, 0.5, 0.75, 1$, respectively. On the other hand, as we discussed in Sec. II, the Bell states are eigenstates of the Kraus operators $E_k^c$ in correlated Pauli channel. As a result, they can pass undisturbed via the full-correlated channel. So when $\mu = 1$, i.e., the channel becomes full-correlated, the capacity of dense coding $\chi$ is kept in 2. Based on above discussion, we can reasonably infer that the capacity of quantum dense coding increases monotonically with the increasing correlation strength $\mu$ when the phase damping coefficient $\lambda$ is fixed. So we can conclude that the correlated channel increases the capacity of quantum dense coding comparing to the uncorrelated channel.

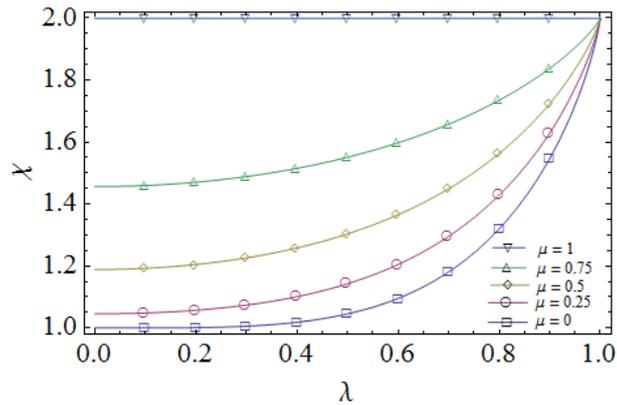

Fig. 3. The dependence of the capacity of quantum dense coding on the phase damping coefficient $\lambda$, where $\alpha = \beta = \sqrt{2}/2$. The blue line with square, purple line with circle, yellow line with rhombus, green line with triangle and light blue line with inverted triangle (from bottom to top) in plot correspond to the correlation strength of channel $\mu = 0, 0.25, 0.5, 0.75, 1$, respectively.

We discussed how weak measurement and reversal measurement affect the capacity of quantum dense coding in phase damped channel as in above section. Besides the normalized factor, the density matrix is

$$\rho_{11} = (-1+n)^2 \alpha^2$$
$$\rho_{44} = (-1+m)^2 \beta^2 \qquad (19)$$
$$\rho_{14} = \rho_{41} = (-1+m)(1-n)\alpha\beta(\lambda^2(-1+\mu)-\mu).$$

According to Eq.(13), the capacity $\chi$ of quantum dense coding with weak measurement and reversal measurement can be obtained analytically, but it's too long to wirte here. Because weak measurement strength $m=1$ and reversal measurement strength $n=1$ is a singular point, we are difficult to find the maximum value of the capacity of quantum dense coding in phase damped channel. For the purpose of simplicity, we use a point to analyze. For example, the capacity with weak measurement is larger than that without weak measurement for $m=0.95$ and $n=0.95$ when the phase damping coefficient $\lambda=0.5$ and correlation parameter $\mu=0.5$. But we find the influence of weak measurement and reversal measurement on the capacity is so weak that we can neglect it.

**Capacity in correlated depolarizing channel**

The depolarizing noise is the quantum operation that depolarizes the state into a completely mixed state. The Kraus operators for a single qubit are $B_k = \sqrt{p_k}\sigma_k$ $(k=0,x,y,z)$, where $p_0 = (1+\lambda)/2$, $p_x = p_y = p_z = (1-\lambda)/6$ and $\lambda = \exp(-\Gamma t)$. Assuming the time interval between channel consecutive applications on the two qubits is very small, the correlated depolarizing channel model can be applicable. At this point, the Kraus operators $E_{i_1 i_2}$ can be written as

$$E_{k_1 k_2} = \sqrt{p_{k_1} p_{k_2}}\sigma_{k_1} \otimes \sigma_{k_2}, \quad (k_1, k_2 = 0, x, y, z). \qquad (20)$$

The Kraus operators $E_{kk}$ are given as

$$E_{kk} = \sqrt{p_k}\sigma_k \otimes \sigma_k, \quad (k_1, k_2 = 0, x, y, z). \qquad (21)$$

According to Eq.(4), the density matrix elements of the two qubits are

$$\rho_{11} = \frac{1}{9}\left(-(2+\lambda)\alpha^2\left(-2+\lambda(-1+\mu)-\mu\right)-(-1+\lambda)\beta^2\left(1+\lambda(-1+\mu)+2\mu\right)\right),$$
$$\rho_{22} = \rho_{33} = \frac{1}{9}\left(-2+\lambda+\lambda^2\right)(-1+\mu),$$
$$\rho_{44} = \frac{1}{9}\left(-(-1+\lambda)\alpha^2\left(1+\lambda(-1+\mu)+2\mu\right)+\beta^2\left((2+\lambda)^2-(-2+\lambda+\lambda^2)\mu\right)\right), \qquad (22)$$
$$\rho_{14} = \rho_{41} = \frac{1}{9}\alpha\beta\left(1-4\lambda(1+\lambda)(-1+\mu)+8\mu\right).$$

Based on Eq. (7), the capacity of quantum dense coding can be obtained analytically, but the expression is too long to write here. The influences of the correlation strength of channel $\mu$ on capacity are depicted in Fig.4. The blue line in Fig.4 corresponds to $\mu=0$, which corresponds to the uncorrelated channel. The purple line with circle, the yellow line with rhombus, the green line with triangle and the light blue line with inverted triangle represent to $\mu=0.25, 0.5, 0.75, 1$, respectively. This correlated noise channel also adapted for the situation in Sec. II, i.e., the Bell states are eigenstates of the Kraus operators $E_k^c$ in correlated Pauli channel. As a result, they can pass undisturbed via the full-correlated channel. So when $\mu=1$, i.e., the channel becomes full-correlated, the capacity of dense coding $\chi$ is kept in 2. Similarly to above part, we can reasonably infer that the capacity of quantum dense coding increases monotonically with the increasing correlation strength $\mu$ when the phase damping coefficient $\lambda$ is fixed. So we can conclude that the correlated channel increases the capacity of quantum dense coding comparing to the uncorrelated channel.

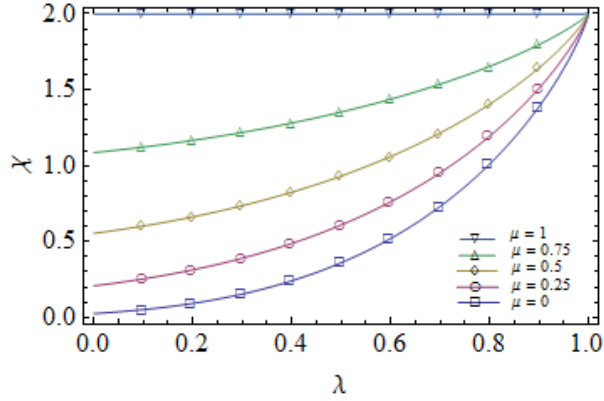

Fig. 4. The dependence of the capacity of quantum dense coding on the depolarizing damping coefficient $\lambda$, where $\alpha=\beta=\sqrt{2}/2$. The blue line with square, purple line with circle, yellow line with rhombus, green line with triangle and light blue line with inversed triangle (from bottom to top) in plot correspond to the correlation strength of channel $\mu=0, 0.25, 0.5, 0.75, 1$, respectively.

Then we discuss the influence of weak measurement and reversal measurement on the capacity.

$$\rho_{11}=(-1+n)^2(-\frac{1}{9}(2+\lambda)\alpha^2(-2+\lambda(-1+\mu)-\mu)-\frac{1}{9}(-1+\lambda)(-1+m)^2\beta^2(1+\lambda(-1+\mu)+2\mu))$$

$$\rho_{22}=\rho_{33}=-\frac{1}{9}(-2+\lambda+\lambda^2)(-1+n)(\alpha^2+(-1+m)^2\beta^2)(-1+\mu)$$

$$\rho_{44}=\frac{1}{9}((1-\lambda)\alpha^2(1+\lambda(-1+\mu)+2\mu)+(-1+m)^2\beta^2((2+\lambda)^2-(-2+\lambda+\lambda^2)\mu))$$

$$\rho_{14}=\rho_{41}=-\frac{1}{9}(1-m)(1-n)\alpha\beta(-1+4\lambda(1+\lambda)(-1+\mu)-8\mu).$$

(23)

We find the capacity with weak measurement is larger than that without weak measurement for $m = 0.99$ and $n = 0.99$ when the depolarizing damping coefficient $\lambda = 0.5$ and correlation strength $\mu = 0.5$. But the influence of weak measurement and reversal measurement on the capacity is very weak whose order is $10^{-2}$.

## IV. CONCLUSION

In this paper, we study the dense coding capacity in correlated noise channels and the influence of weak measurement and reversal measurement on the capacity. We find two conclusions: the first one is correlated noisy channel can improve the capacity of dense coding than in uncorrelated noisy channel. It's easy to understand physically. Because the correlated noisy channel is the situation where time correlations can't neglect. It means two qubits sent almost at the same time, the channel's properties will be unchanged. But time lapse is larger in uncorrelated noisy channel. So the capacity of the correlated noisy channel is larger than that in uncorrelated noisy channel. The second conclusion is weak measurement and reversal measurement can make further efforts on improving the capacity of dense coding in correlated amplitude damping channel, but this improving effort is very small in correlated phase damping channel and correlated depolarizing channel. The explanations about this are correlated amplitude damping channel is non-Pauli channel, but correlated phase damping channel and correlated depolarizing channel are Pauli channel, so the improvements by weak measurement and reversal measurement are different.

**Acknowledgements**

This work was supported by the National Natural Science Foundation of China (Grant No. 12074027).